\documentclass[lettersize,journal]{IEEEtran}

\usepackage{amsmath,amsfonts}
\usepackage{algorithmic}
\usepackage{array}

\usepackage[caption=false,font=footnotesize]{subfig}

\usepackage{textcomp}
\usepackage{stfloats}
\usepackage{url}
\usepackage{verbatim}
\usepackage{graphicx}

\usepackage{makecell}
\usepackage{multirow}
\usepackage{booktabs}

\usepackage[table]{xcolor}
\usepackage{pifont}
\usepackage{bm}

\usepackage[citecolor=blue,colorlinks]{hyperref}
\usepackage{fancyhdr}
\usepackage{balance}

\def\BibTeX{{\rm B\kern-.05em{\sc i\kern-.025em b}\kern-.08em
    T\kern-.1667em\lower.7ex\hbox{E}\kern-.125emX}}
\begin{document}
\title{LC3EM: Long-Range Context Extrapolation Enhanced Entropy Model for Coordinate-based Overfitting Image Codecs}

\author{Hao Wang, Junyan~Huo,~\IEEEmembership{Member,~IEEE,} 
Fei~Yang,
Shuai~Wan,~\IEEEmembership{Member,~IEEE,}
Fuzheng~Yang,~\IEEEmembership{Member,~IEEE}

\thanks{This work has been submitted to the IEEE for possible publication.
Copyright may be transferred without notice, after which this version may no longer be accessible.}

\thanks{Hao Wang, Junyan Huo and Fuzheng Yang are with the School of Telecommunication Engineering, Xidian University, Xi’an 710071, China (e-mail: 18010190023@stu.xidian.edu.cn; jyhuo@mail.xidian.edu.cn; fzhyang@mail.xidian.edu.cn). Fuzheng Yang is the corresponding author.}
\thanks{Fei Yang is with the College of Computer Science, Nankai University, Tianjin, China (e-mail: feiyang@nankai.edu.cn).}
\thanks{Shuai Wan is with the School of Electronics and Information, Northwestern Polytechnical University, Xi’an, China, and also with the School of Engineering, Royal Melbourne Institute of Technology, Melbourne, VIC 3001, Australia (e-mail: swan@nwpu.edu.cn).}}

\maketitle

\begin{abstract}

Coordinate-based overfitting image codecs have attracted increasing attention for their low decoding complexity and independence from cross-image generalization. However, representative approaches such as COOL-CHIC face an inherent entropy-modeling trade-off: lightweight models have limited capacity, while more expressive ones incur additional bitrate overhead from transmitting image-specific parameters. Inspired by the prediction mechanism in traditional codecs, we propose a new entropy-modeling strategy that introduces complementary prediction modes with region-adaptive soft mode selection, rather than relying on a single learned predictor to model diverse types of redundancy. Based on this concept, we develop a Long-Range Context Extrapolation Enhanced Entropy Model (LC3EM), which can be readily integrated into coordinate-based overfitting codecs. Specifically, a parameter-free Neighborhood-based Linear Extrapolation Mode (NLEM) complements the tiny MLP-based local predictor to exploit long-range contextual redundancy and strongly directional structures. A Minimum-Entropy-Inspired Continuous Mode Selection strategy is designed to adaptively fuse these two complementary modes, while requiring the transmission of only the parameters of a single additional linear layer. Moreover, to alleviate the mismatch between training-time relaxed and actual discrete quantization, we introduce a lightweight iterative latent rounding refinement stage to further improve compression performance. Extensive experiments demonstrate consistent improvements across diverse benchmarks, particularly on highly regular computer-generated images. When integrated with COOL-CHIC 4.0, the proposed method achieves BD-rate gains of -3.43\% and -7.69\% on the SIQAD and API datasets, respectively. With COOL-CHIC 5.0 as the backbone, the corresponding gains are -2.88\% and -3.15\%, respectively. The code will be made publicly available soon at \url{https://github.com/xduwh321/LC3EM_overfitting_image_compression}.

\end{abstract}


\section{Introduction}
\IEEEPARstart{A}{s} a fundamental medium for multimedia communication, images require effective compression algorithms for transmission and storage. In recent years, learned image compression (LIC), particularly autoencoder-based approaches~\cite{balle2016end,balle2018variational,minnen2018joint,minnen2020channel}, has attracted considerable attention due to its strong rate-distortion (RD) performance. While these methods have achieved impressive compression performance, several practical challenges remain to be addressed, including relatively high decoding complexity~\cite{huang2024unveiling} and limited generalization to out-of-domain images~\cite{shen2023dec}.

\begin{figure}[t]
  \centering
  \includegraphics[width=1\linewidth]{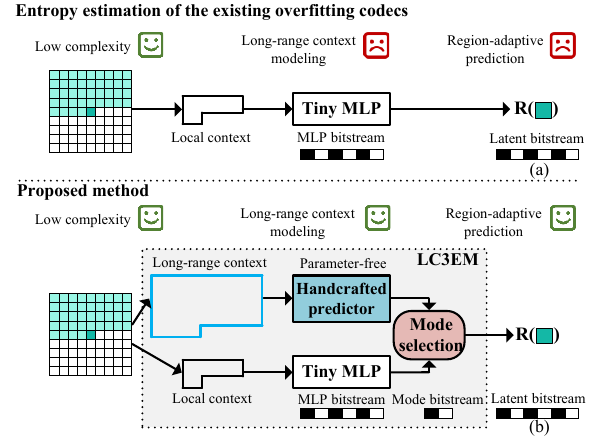}
  \caption{Comparison of the latent entropy coding processes in advanced coordinate-based overfitting image codecs and the proposed strategy. R(*) denotes the coding cost of the latent at the corresponding position.}
  \label{fengmiantu}
\end{figure}

More recently, per-image overfitting compression has emerged as a promising alternative paradigm. Instead of learning a single model that generalizes across diverse images, these methods optimize a compact model for each image independently. This formulation can help alleviate the aforementioned challenges, as decoding can remain lightweight while the dependence on cross-image generalization is substantially reduced. The seminal work COIN~\cite{dupont2021coin} pioneered this paradigm by representing an image with neural network parameters, which are transmitted to the decoder and map pixel coordinates to corresponding RGB values. While demonstrating the potential of per-image neural representation for compression, its rate-distortion performance remained relatively modest and mainly surpassed JPEG~\cite{wallace2002jpeg} at low bitrates. A major breakthrough came with COOL-CHIC~\cite{ladune2023cool} (Coordinate-based Low Complexity Hierarchical Image Codec), which represents an image as a set of latent grids, reconstructs it with a compact synthesis network, and compresses the latent grids using a lightweight entropy model. With this hybrid approach, COOL-CHIC significantly improves compression efficiency and achieves RD performance comparable to H.265/HEVC~\cite{sullivan2012overview}. Subsequent works~\cite{blard2024overfitted,leguay2023low,kim2024c3} further enhanced this paradigm through architectural refinements and improved optimization strategies. These advances were subsequently consolidated in COOL-CHIC 4.0, which delivers strong RD performance and is competitive with H.266/VVC~\cite{ohm2018versatile}. More recently, COOL-CHIC 5.0~\cite{ladune2026cool} further improved this approach, achieving RD performance comparable to advanced autoencoder-based codecs and establishing state-of-the-art performance among overfitted codecs. Through these successive developments, COOL-CHIC has become a representative framework for overfitting-based image compression.

As the framework has matured, further improving the RD performance of coordinate-based overfitting codecs remains challenging. A major reason lies in the trade-off imposed by the entropy model adopted in this framework. On the one hand, \textit{its highly compact design limits the modeling capacity of the learned local predictor, making it difficult to capture long-range contextual information}, which in turn constrains its ability to fully exploit redundancies in the latent grids. On the other hand, \textit{the entropy model cannot be expanded arbitrarily}, since all networks involved must be transmitted as part of the bitstream, and any increase in model capacity directly results in additional bit overhead, as shown in Figure~\ref{fengmiantu}(a).

This dilemma naturally raises the following question: \textit{Can the capability of the entropy model be enhanced with negligible additional overhead for transmitting network parameters?} Traditional handcrafted image codecs offer valuable inspiration through the multiple-hypothesis prediction paradigm~\cite{multihypo}, where complementary prediction modes are designed for diverse texture and structural characteristics. During encoding, rate-distortion optimization (RDO) selects the suitable mode for each coding unit, while only a compact mode index is signaled to the decoder, enabling content-adaptive prediction with negligible overhead. Motivated by this principle, we address the entropy-modeling dilemma in coordinate-based overfitting codecs by introducing an additional handcrafted prediction mode and enabling region-adaptive fusion through soft mode selection. The core concept is illustrated in Figure~\ref{fengmiantu}(b).

Based on this concept, the Long-Range Context Extrapolation Enhanced Entropy Model (LC3EM) is proposed in this paper, providing a general approach for enhancing entropy modeling in coordinate-based overfitting codecs. Specifically, an additional Neighborhood-based Linear Extrapolation Mode (NLEM) is introduced to capture texture patterns from long-range  contexts. This mode establishes a prediction process that integrates long-range contextual sample construction, template matching, and least-squares estimation. As a handcrafted module, NLEM introduces no additional transmission overhead and effectively complements the tiny MLP-based local predictor, particularly excelling at modeling edges and directional
structural redundancy. Building upon these two predictors, a Minimum-Entropy-Inspired Continuous Mode Selection strategy, termed MECMS, is introduced. This strategy adapts the traditional mode selection mechanism to a networked setting, upgrading simple prediction weighting to distribution-level fusion motivated by an entropy-reduction perspective, thereby enabling the two modes to complement each other and fully exploit their respective strengths while requiring the transmission of only the parameters of a single additional linear layer. Furthermore, since the proposed framework, particularly its handcrafted NLEM, is highly sensitive to noise, we develop an iterative latent rounding refinement stage inspired by the stochastic annealing-based methods in~\cite{yang2020improving,perugachi2024robustly}. This stage mitigates the discretization gap between noise-relaxed training and deterministic rounding in practical coding, thereby further improving the rate-distortion performance.

Our contributions are summarized as follows:

\begin{enumerate}
\item A new entropy modeling strategy is proposed by introducing complementary prediction modes and region-adaptive soft mode selection. Based on this concept, we develop a Long-Range Context Extrapolation Enhanced Entropy Model (LC3EM) to enhance the entropy modeling capability of coordinate-based overfitting image codecs, thereby improving their RD performance.

\item In LC3EM, a Neighborhood-based Linear Extrapolation Mode (NLEM) is introduced to complement the MLP-based local predictor by capturing long-range contextual and directional structural redundancy. A Minimum-Entropy-Inspired Continuous Mode Selection (MECMS) strategy is developed to adaptively integrate the two complementary predictors from a distributional perspective, thereby improving entropy estimation.

\item To mitigate the adverse effect of relaxation noise on the proposed method, we develop an iterative latent rounding refinement stage during training to further improve the rate-distortion performance.

\item Experiments on the COOL-CHIC series across benchmark datasets covering diverse content types demonstrate the effectiveness of the proposed method, with particularly pronounced gains on highly regular computer-generated images.

\end{enumerate}

The remainder of this paper is organized as follows. Section~\ref{sec2} reviews recent related work on image compression methods. Section~\ref{sec:background} introduces the background of COOL-CHIC. Section~\ref{sec4} presents the proposed method in detail. Experimental results are reported in Section~\ref{sec5}. Finally, Section~\ref{sec6} concludes the paper.

\section{Related work}
\label{sec2}
\subsection{VAE-based learned image compression}
Variational autoencoder~\cite{kingma2013auto} (VAE)-based learned image compression was first introduced by Ball\'e \emph{et al.}~\cite{balle2016end}, showing that optimizing the evidence lower bound (ELBO) aligns closely with the RD trade-off. Subsequent works~\cite{balle2018variational,minnen2018joint,minnen2020channel} improved entropy models for better latent distribution estimation, establishing the canonical VAE-based compression pipeline: an analysis transform maps an input to latent features, a synthesis transform reconstructs the image, and an entropy model with hyperpriors and context models performs entropy estimation. Later studies enhanced transforms and entropy models using residual blocks~\cite{he2022elic}, attention~\cite{cheng2020learned,liu2019non}, Transformers~\cite{liu2023learned,qian2022entroformer}, Mamba~\cite{qin2024mambavc,zeng2025mambaic}, and RWKV~\cite{feng2025linear}, progressively narrowing the gap to, and in some cases substantially outperforming, traditional standards such as H.266/VVC~\cite{ohm2018versatile}. While VAE-based methods have achieved notable success, they still suffer from high decoding complexity~\cite{huang2024unveiling} and limited generalization to out-of-domain data~\cite{shen2023dec}, which restrict their applicability in practical deployment scenarios.

\subsection{Coordinate-based overfitting image compression}

In recent years, coordinate-based overfitting image compression has emerged as an alternative, where a model is optimized for each individual image and the resulting parameters are compressed. COIN~\cite{dupont2021coin} pioneered this approach by encoding a single image into a neural network and compressing it via parameter quantization and entropy coding. COIN++~\cite{dupont2022coin++} extended the framework with feature modulation and meta-learning, though its RD performance remained below that of advanced codecs. COOL-CHIC~\cite{ladune2023cool} introduced a hybrid representation combining coordinate-based latent grids with lightweight MLPs, significantly improving RD performance and narrowing the gap with mainstream compression methods. Subsequent refinements~\cite{leguay2023low} produced the COOL-CHIC~2.x series, while components of C3~\cite{kim2024c3} enhanced noise injection by replacing uniform noise with soft quantization~\cite{agustsson2020universally} and Kumaraswamy noise~\cite{kumaraswamy1980generalized}, yielding substantial gains and making COOL-CHIC competitive with H.266/VVC on benchmarks such as Kodak~\cite{franzen_kodak_truecolor}. Further improvements~\cite{blard2024overfitted} optimized the decoding process, with these advances consolidated in the official COOL-CHIC~4.0 implementation. After the release of COOL-CHIC~4.0, several studies~\cite{limoric,wu2025lotterycodec,benjak2026lance} further explored the architectural design space of overfitting-based image compression networks. In particular, LANCE~\cite{benjak2026lance}, building upon COOL-CHIC~4.0,
introduced locally adaptive context estimation through a forward-signaled
spatial hyperprior, allowing the entropy model to adapt to spatially varying
image statistics and further improving RD performance.  More recently, COOL-CHIC~5.0 introduced inter-feature context modeling, linear stabilizers, the SOAP~\cite{vyas2025soap} optimizer, and refined optimization strategies, collectively establishing a new state-of-the-art in overfitting-based image compression. It substantially outperforms H.266/VVC on high-resolution images while achieving RD performance competitive with advanced autoencoder-based codecs. Beyond improving RD performance, recent efforts have extended overfitting-based compression to video~\cite{leguay2024cool,leguay2025improved}, accelerated per-instance optimization~\cite{zhang2025mliic,borrell2026hypercool}, and improved perceptual quality~\cite{philippeperceptually,balle2024good,ladune2024cool}. Despite extensive research efforts on coordinate-based overfitting codecs, the capacity limitation of entropy models imposed by parameter transmission overhead remains largely unexplored, hindering further performance improvements.

\subsection{Prediction in traditional codecs}

Images exhibit regular and predictable structures, which can be regarded as redundancy from a compression perspective. Prediction in traditional codecs fundamentally exploits such redundancy to reduce the coding rate. Due to the diversity of these structures, however, a single handcrafted prediction model cannot effectively handle all scenarios. Therefore, traditional codecs employ a set of complementary prediction modes, each designed to capture specific types of redundancy. For instance, H.266/VVC~\cite{ohm2018versatile} supports 67 intra prediction modes, together with additional prediction tools~\cite{pfaff2021intra}. During encoding, rate-distortion optimization (RDO) selects the most suitable mode for each coding unit, while the decoder reproduces the corresponding prediction according to the signaled mode information. For coordinate-based overfitting image compression networks, the capacity of entropy models is constrained by the parameter transmission overhead, making them insufficiently powerful to capture diverse contextual dependencies. From this perspective, introducing additional parameter-free modes that complement the tiny MLP offers a promising way to further improve performance. Inspired by context-based prediction methods~\cite{delp1979image,chellappa1985texture,kokaram2004statistical,xu2025intra}, NLEM is designed to compensate for the limited long-range context modeling capability of the COOL-CHIC entropy model.

\section{Overview of COOL-CHIC}
\label{sec:background}

In this section, we briefly review COOL-CHIC~\cite{ladune2023cool}, as it is the predominant state-of-the-art framework for overfitting-based image compression. COOL-CHIC consists of three parts: a multi-resolution coordinate-based latent grid $I=(I^1,I^2,\ldots,I^L)$, a synthesis network $f_{\theta}$ for image reconstruction, and a context network $g_c$ for entropy modeling.

\noindent\textbf{Coordinate-based latent grid:} The multi-resolution latent grid represents the image in a coarse-to-fine manner, with spatial resolutions $((h,w),(h/2,w/2),\ldots,(h/2^{L-1},w/2^{L-1}))$, where $h$ and $w$ are the height and width of the original image. Higher-resolution grids capture fine details, whereas lower-resolution grids represent coarse structural information. In COOL-CHIC 4.0, $L=7$, whereas COOL-CHIC 5.0~\cite{ladune2026cool} introduces three additional hyperlatents.

\noindent\textbf{Synthesis:} The synthesis network $f_{\theta}$ reconstructs the image from the multi-resolution latent grids. Specifically, grids at different resolutions are first upsampled to the highest resolution and then fed into a lightweight network for reconstruction.

\noindent\textbf{Entropy model:} The entropy model employs a context network $g_c$ to capture redundancy in the multi-resolution latent representation. Each latent feature is modeled by a Laplace distribution with parameters $(\mu, b)$. Let $p=(x,y)$ denote a target location with feature value $I(p)$:
\begin{equation}
I(p) \sim \mathcal{L}\!\bigl(\mu(p),\, b(p)\bigr),
\end{equation}
\begin{equation}
\mu(p), b(p) = g_c\!\bigl(\mathrm{context}(p)\bigr),
\end{equation}
where $\mathrm{context}(p)$ denotes the previously decoded context at location $p$. Under the High Operating Point (HOP) configuration, $g_c$ in COOL-CHIC 4.0 takes 16 same-resolution context elements as input for entropy estimation. In COOL-CHIC 5.0, the context is extended to 20 elements, comprising 14 same-resolution contexts and 6 lower-resolution contexts generated from already decoded elements through an additional linear layer. These contexts are concatenated and fed into $g_c$ for entropy estimation.

\begin{figure*}[t]
  \centering
  \includegraphics[width=0.98\linewidth]{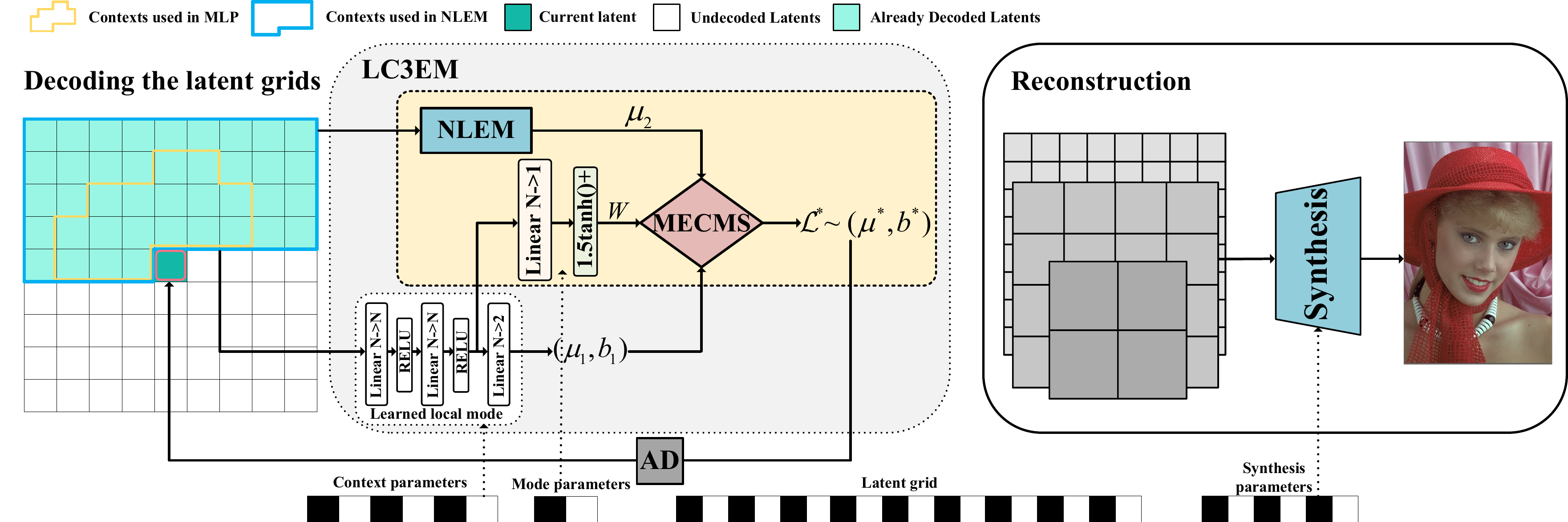}
  \caption{Overview of the decoding pipeline of the proposed method. The left part illustrates the decoding process of the latent grids, while the right part shows the reconstruction process. The overall architecture of the proposed LC3EM is enclosed in the gray box, and the additional modules introduced beyond the original COOL-CHIC are highlighted in the yellow box. AD denotes arithmetic decoding. The example shown in the figure illustrates the proposed extension based on the COOL-CHIC 4.0 backbone. When using the COOL-CHIC 5.0 backbone, the only difference lies in  context numbers and how they are incorporated in the learned local mode.}
  \label{fig:coolchic_all}
\end{figure*}

\section{Proposed method}
\label{sec4}

In this paper, we propose a Long-Range Context Extrapolation Enhanced Entropy Model (LC3EM) to improve the entropy modeling capability of the COOL-CHIC series. The overall enhanced architecture is illustrated in Figure~\ref{fig:coolchic_all}. Specifically, LC3EM introduces two additional components, namely the Neighborhood-based Linear Extrapolation Mode (NLEM) and the Minimum-Entropy-Inspired Continuous Mode Selection (MECMS), as highlighted by the yellow box in the figure. These two components will be introduced in detail in Sections~\ref{nlem} and~\ref{nmmmecms}, respectively. Furthermore, we propose an iterative latent rounding refinement stage to further improve the rate–distortion performance, which is described in detail in Section~\ref{sga}.

\subsection{Neighborhood-based linear extrapolation mode}
\label{nlem}

For a target location \(p\) in latent grids, the value \(I(p)\) is predicted from already decoded neighboring features using a linear model in NLEM. Let \(\mathbf{x}_p \in \mathbb{R}^{c \times 1}\) denote the neighborhood feature vector at \(p\), and let \(\mathbf{a} \in \mathbb{R}^{c \times 1}\) be the extrapolation coefficients. The prediction is given by

\begin{equation}
\hat{I}(p) = \mathbf{x}_p^{\mathsf{T}} \mathbf{a}.
\end{equation}
The NLEM consists of three steps: (i) long-range context sample construction, (ii) extrapolation coefficient estimation, and (iii) feature prediction.

\noindent\textbf{Step 1: long-range context sample construction.}
First, a set of \(s\) already decoded context locations in the vicinity of \(p\) is collected,
\begin{equation}
C_p = \{q_1, q_2, \ldots, q_s\}, \qquad (s > c).
\end{equation}
For each context position \(q_k\), the neighborhood feature vector \(\mathbf{x}_{q_k} \in \mathbb{R}^{c \times 1}\) is formed following the same rule used at \(p\). These transposed vectors are stacked row-wise to obtain the observation matrix
\begin{equation}
\mathbf{X} =
\begin{bmatrix}
\mathbf{x}_{q_1} & \mathbf{x}_{q_2} & \cdots & \mathbf{x}_{q_s}
\end{bmatrix}^{\mathsf{T}}
\in \mathbb{R}^{s \times c},
\end{equation}
with the corresponding observation vector
\begin{equation}
\mathbf{y} =
\begin{bmatrix}
I(q_1) & I(q_2) & \cdots & I(q_s)
\end{bmatrix}^{\mathsf{T}}
\in \mathbb{R}^{s \times 1}.
\end{equation}

In principle, \(s\) and \(c\) can be chosen arbitrarily, as long as the matrix has full column rank and \(s > c\). This allows NLEM to exploit richer long-range contextual information, at the cost of additional computational complexity. In this paper, we set \(s = 40\) and \(c = 4\), enabling NLEM to utilize more global context than the MLP-based model (which uses $\leq 20$ context samples) while keeping the complexity manageable. Furthermore, not all context samples contribute equally to predicting \(p\). To emphasize samples that are more representative of the local structure around \(p\), each candidate \(q_k\) is assigned a nonnegative similarity score \(S_k\) based on the distance between local templates. Let \(\mathbf{t}(p) \in \mathbb{R}^{d_t \times 1}\) be a local template descriptor at \(p\), and \(\mathbf{t}(q_k) \in \mathbb{R}^{d_t \times 1}\) the corresponding descriptor at \(q_k\). In this paper, the template descriptor is constructed using the same neighborhood sampling operation as \(\mathbf{x}_{q_k}\). A similarity score is defined as

\begin{equation}
S_k = \exp\left(
-\frac{\|\mathbf{t}(q_k)-\mathbf{t}(p)\|_2^2}
{2d_t\rho^2}
\right),
\qquad S_k \in (0, 1],
\end{equation}

where \(\rho = 0.1\) controls the decay. To avoid extreme values that may lead to ill-conditioned estimation, \(S_k\) is clipped to a bounded range, e.g., \(S_k \in [S_{\min}, 1]\) with \(S_{\min} = 10^{-3}\) in our implementation. The diagonal similarity matrix is defined as
\begin{equation}
\mathbf{S} = \mathrm{diag}(S_1, \ldots, S_s).
\end{equation}

\noindent\textbf{Step 2: Extrapolation coefficient estimation.}
Given $(\mathbf{X}, \mathbf{y}, \mathbf{S})$, the extrapolation coefficients $\mathbf{a}$ are estimated via weighted least squares (WLS):
\begin{equation}
\mathbf{a}^{\ast}
=
\arg\min_{\mathbf{a}}
\left\lVert
\mathbf{S}^{1/2}
\left(
\mathbf{y}-\mathbf{X}\mathbf{a}
\right)
\right\rVert_2^2,
\end{equation}
which leads to the weighted normal equations
\begin{equation}
\mathbf{X}^{\mathsf{T}}\mathbf{S}\mathbf{X}\mathbf{a}
=
\mathbf{X}^{\mathsf{T}}\mathbf{S}\mathbf{y}.
\end{equation}

\begin{figure}[t]
  \centering
  \includegraphics[width=1.0\linewidth]{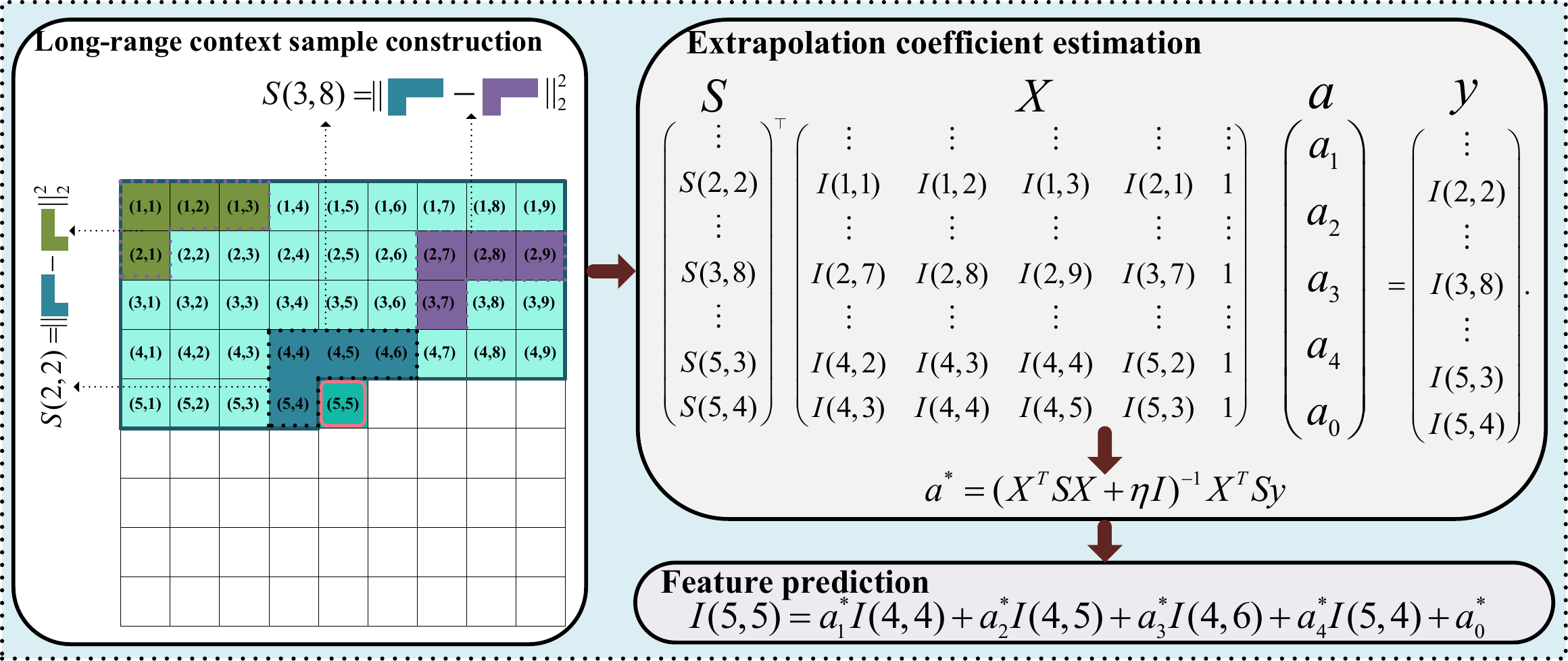}
  \caption{Example of the proposed NLEM for predicting $I(5,5)$.}
  \label{fig:nlem}
\end{figure}

To improve numerical stability, ridge regularization is introduced, yielding
\begin{equation}
\left(
\mathbf{X}^{\mathsf{T}}\mathbf{S}\mathbf{X}
+
\eta\mathbf{I}_c
\right)\mathbf{a}
=
\mathbf{X}^{\mathsf{T}}\mathbf{S}\mathbf{y},
\end{equation}
where $\mathbf{I}_c$ denotes the identity matrix.

Rather than using a fixed regularization strength, $\eta$ is adaptively determined according to the local scale of the weighted normal matrix:
\begin{equation}
\eta
=
\epsilon
\cdot
\frac{1}{c}
\mathrm{tr}
\left(
\mathbf{X}^{\mathsf{T}}\mathbf{S}\mathbf{X}
\right),
\end{equation}
where $\epsilon=0.01$ and $\mathrm{tr}(\cdot)$ denotes the matrix trace. Accordingly, the optimal extrapolation coefficients are obtained as
\begin{equation}
\mathbf{a}^{\ast}
=
\left(
\mathbf{X}^{\mathsf{T}}\mathbf{S}\mathbf{X}
+
\eta\mathbf{I}_c
\right)^{-1}
\mathbf{X}^{\mathsf{T}}\mathbf{S}\mathbf{y}.
\end{equation}

\noindent\textbf{Step 3: feature prediction.}
With the estimated coefficients \(\mathbf{a}^{\ast}\), the feature at \(p\) is predicted by
\begin{equation}
\hat{I}(p) = \mathbf{x}_p^{\mathsf{T}} \mathbf{a}^{\ast}.
\end{equation}

The above procedure is applied at each spatial location to perform point-wise prediction. Compared with the learned local mode in COOL-CHIC, NLEM requires no additional parameter transmission, while enabling the use of long-range context and effectively exploiting spatial redundancies with strong directional characteristics. For clarity, Figure~\ref{fig:nlem} illustrates the overall procedure, using the prediction at location \(p = (5,5)\) as an example. In our practical implementation, an additional constant term \(a_0\) is also estimated and incorporated; we omit it from the above formulation for simplicity. By linearly extrapolating the current value from the contextual texture, NLEM can effectively exploit long-range and even global redundancy, providing a powerful prediction mechanism that achieves particularly high accuracy for highly regular and strongly directional structures.

\subsection{Minimum-entropy-inspired continuous mode selection}
\label{nmmmecms}

As mentioned above, the MLP-based learned local mode and the NLEM exhibit complementary behaviors under different spatial contexts. In traditional codecs such as H.266/VVC, such complementarity is typically exploited through discrete mode selection. However, in COOL-CHIC, point-wise entropy modeling makes the transmission of per-latent mode indices impractical, while the non-differentiability of hard selection is incompatible with end-to-end training. To overcome these limitations, a differentiable fusion method from a distributional perspective, termed MECMS, is introduced as a continuous surrogate for mode selection.

Let \(Z_1\) and \(Z_2\) denote the predictive random variables associated with the learned local mode and NLEM, respectively, with location-dependent means \(\mu_1\) and \(\mu_2\) and prediction variances \(\sigma_1^2\) and \(\sigma_2^2\). We first consider the following continuous fusion in the random-variable domain:
\begin{equation}
Z^\ast = w Z_1 + (1-w) Z_2, \qquad w \in \mathbb{R}.
\label{eq:rv_fusion}
\end{equation}

\begin{figure}[h]
  \centering
  \includegraphics[width=0.8\linewidth]{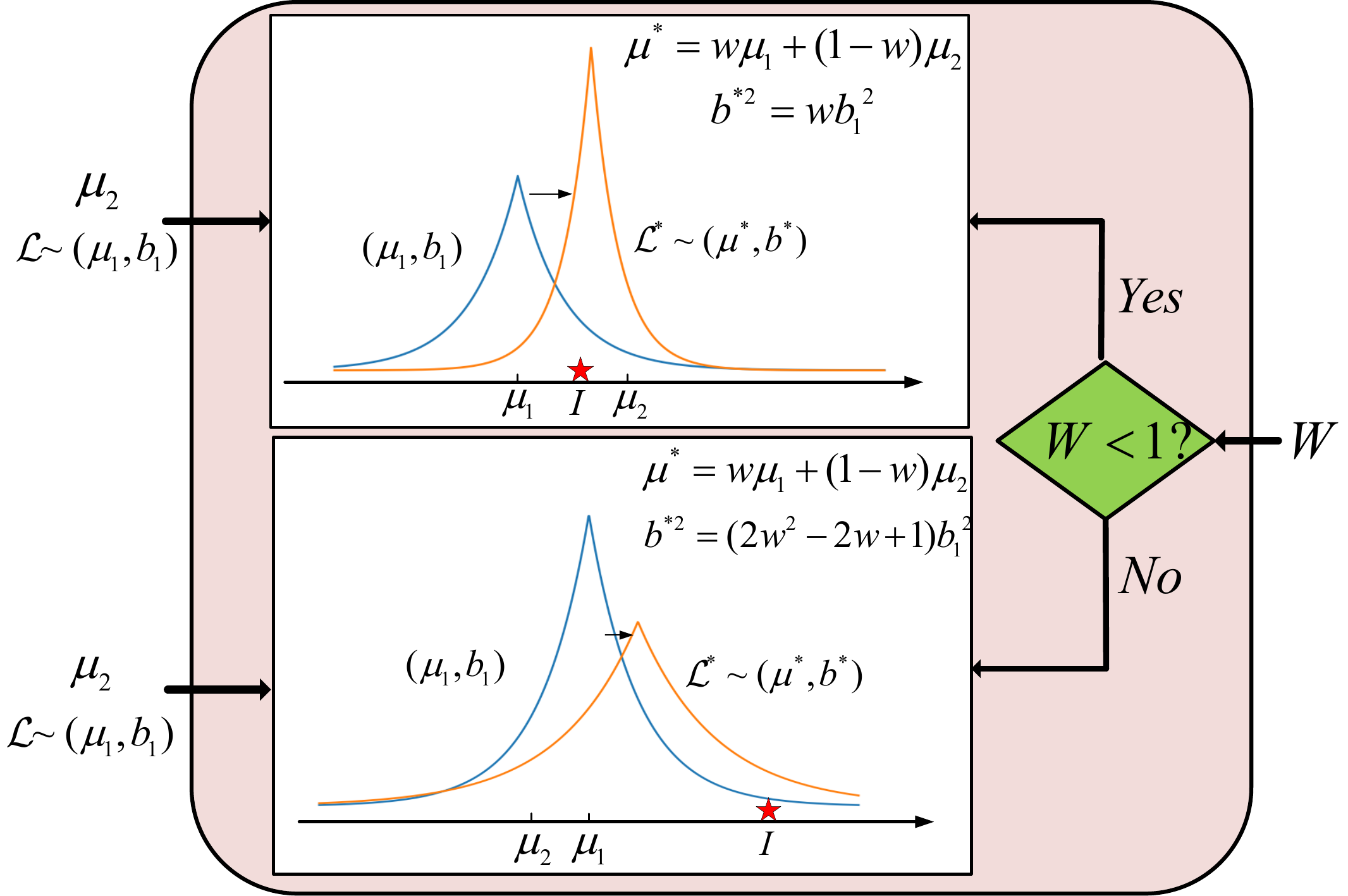}

  \caption{Architecture of the proposed MECMS. The blue and yellow curves correspond to the distributions before and after fusion, respectively.}
  \label{mecms}
\end{figure}

The design of MECMS is motivated by an entropy-reduction perspective. For a fixed location-scale distribution family, such as Gaussian or Laplace distributions, the differential entropy is a monotonic function of the distribution scale, and hence of its variance. Therefore, variance reduction provides a surrogate for analyzing entropy reduction. To obtain a simple
analytical interpretation, we consider the idealized case in which
$Z_1$ and $Z_2$ are approximately uncorrelated. Under this assumption, the variance of the fused random variable in \eqref{eq:rv_fusion} is
\begin{align}
\mathrm{Var}(Z^\ast)
&= \mathrm{Var}\!\left(w Z_1 + (1-w) Z_2\right) \notag \\
&= w^2 \sigma_1^2 + (1-w)^2 \sigma_2^2.
\label{eq:var_fusion}
\end{align}

Minimizing \eqref{eq:var_fusion} with respect to \(w\) yields
\begin{equation}
w^\ast = \frac{\sigma_2^2}{\sigma_1^2 + \sigma_2^2},
\label{eq:w_opt_mu}
\end{equation}
where \(w^\ast\) is the minimum-variance fusion weight. Substituting \(w^\ast\) into \eqref{eq:rv_fusion} gives
\begin{equation}
\begin{aligned}
\mu^\ast
&= w^\ast \mu_1 + (1-w^\ast)\mu_2, \\
(\sigma^\ast)^2
&= \frac{\sigma_1^2 \sigma_2^2}{\sigma_1^2 + \sigma_2^2}
= w^\ast \sigma_1^2
= (1-w^\ast)\sigma_2^2.
\end{aligned}
\label{eq:mu_sigma_fusion}
\end{equation}

As indicated by \eqref{eq:mu_sigma_fusion}, this formulation provides a useful theoretical interpretation of entropy-reducing fusion. In particular, when a new predictor is incorporated into a baseline predictor, an effective fusion weight should not only refine the mean estimate but also reduce the uncertainty associated with the baseline prediction. Moreover, the magnitude of this uncertainty reduction should be directly controlled by the fusion weight itself.

In practice, we directly parameterize the fused entropy model as a Laplace distribution, $Z^\ast \sim \mathcal{L}(\mu^\ast,b^\ast)$, to remain compatible with the original COOL-CHIC framework, where the variance of a Laplace distribution is $2(b^\ast)^2$. Since NLEM is a deterministic handcrafted predictor and does not provide an explicit uncertainty estimate, the closed-form solution in \eqref{eq:w_opt_mu} cannot be directly evaluated. Instead, inspired by the minimum-variance relation in \eqref{eq:mu_sigma_fusion}, we introduce a learnable fusion weight \(w\) that jointly controls the interpolation of the prediction mean and the modulation of the scale parameter:
\begin{equation}
\mu^\ast = w \mu_1 + (1-w)\mu_2,
\qquad
(b^\ast)^2 = w\, b_1^2,
\label{eq:laplace_fusion}
\end{equation}
where \(\mu_1\) and \(b_1\) are predicted by the learned local mode in COOL-CHIC, and \(\mu_2\) is provided by NLEM.  Equation~\eqref{eq:laplace_fusion} provides a practical parameterization motivated by the minimum-entropy analysis above.  In this formulation, a smaller \(w\) assigns greater importance to NLEM while simultaneously reducing the scale of the fused distribution, reflecting increased confidence in the complementary prediction.  

The formulation above constitutes the core of MECMS for fusing the two modes at the distribution level, as illustrated in the upper part of Figure~\ref{mecms}.
Learning only one additional weight \(w \in [0,1]\) is sufficient to achieve effective fusion between the two predictors. Nevertheless, this formulation is most effective when the true latent value lies between the two predictions. Special cases are also considered, in which the target feature cannot be well predicted by either predictor. In such cases, blindly reducing the scale may instead degrade coding performance. To address this issue, the range of \(w\) is extended from \([0,1]\) to \([0,1.5]\), and different fusion strategies are adopted over the intervals \([0,1]\) and \((1,1.5]\).

In practice, MECMS first introduces a lightweight auxiliary network 
$\mathrm{g}_w$, consisting of only a single fully connected layer, followed by a 
$\tanh$ activation, to predict the fusion weight:
\begin{equation}
w = \bigl(1.5\,\tanh(\mathrm{g}_w(\mathbf{h}))\bigr)_{+},
\label{eq:w_predict}
\end{equation}
where $\mathbf{h}$ denotes the penultimate hidden representation of the baseline 
MLP, and $(\cdot)_+$ denotes clamping at zero. For numerical stability, $w$ is 
further lower-bounded by $10^{-8}$ in implementation to prevent the fused scale 
parameter from degenerating to zero.

When \(w \in [0,1]\), the fusion follows the entropy-reduction-inspired parameterization in \eqref{eq:laplace_fusion}. When \(w \in (1,1.5]\), a special case is triggered, as illustrated in the lower part of Figure~\ref{mecms}. In this regime, the fusion of the mean is allowed to move beyond the interval spanned by \(\mu_1\) and \(\mu_2\). Accordingly, the update rule for the mean remains unchanged, whereas the scale parameter \(b^\ast\) is updated as
\begin{equation}
(b^\ast)^2 = \left(2w^2 - 2w + 1\right)b_1^2.
\label{eq:w_gt_1}
\end{equation}
The derivation of \eqref{eq:w_gt_1} additionally assumes that NLEM and the MLP-based mode have the same prediction variance, and then substitutes this assumption into \eqref{eq:var_fusion}.


\subsection{Iterative latent rounding refinement stage}
\label{sga}

During optimization, the COOL-CHIC series replaces discrete quantization with a noise-relaxed approximation, whereas practical entropy coding relies on deterministically rounded integer latents. This mismatch may degrade the final coding performance and is particularly relevant to LC3EM, since NLEM estimates its extrapolation coefficients from previously decoded causal latents. Small perturbations in these latents may alter the neighborhood values and consequently the corresponding least-squares solution, thereby affecting the distribution estimation of subsequent latents.

To alleviate this issue, we introduce an iterative latent rounding refinement stage after the standard COOL-CHIC optimization. Specifically, we adopt the two-class SGA (Stochastic Gumbel Annealing) framework of Yang \textit{et al.}~\cite{yang2020improving} together with the sigmoid scaled logit (SSL) relaxation proposed by Perugachi-Diaz \textit{et al.}~\cite{perugachi2024robustly}. During refinement, all network parameters are frozen and only the latent grids \(I\) are updated. We further adapt the refinement procedure to the proposed method through a two-stage rate estimation strategy.

For a latent element \(I(p)\), its two neighboring integer candidates are defined as
\begin{equation}
I^{-}(p)=\lfloor I(p)\rfloor,\qquad
I^{+}(p)=I^{-}(p)+1.
\label{eq:round_candidates}
\end{equation}
The corresponding distances are
\begin{equation}
d^{-}(p)=I(p)-I^{-}(p),\qquad
d^{+}(p)=1-d^{-}(p).
\end{equation}
Following the two-class SSL formulation in~\cite{perugachi2024robustly}, the score of each rounding candidate is defined as
\begin{equation}
\ell_r(p)
=
-\log\left[
1+
\left(
\frac{d^r(p)}{1-d^r(p)}
\right)^{\alpha}
\right],
\qquad r\in\{-,+\},
\label{eq:ssl_score}
\end{equation}
where \(\alpha\) controls the rounding preference.

To construct a differentiable stochastic relaxation, independently sampled Gumbel noise \(\xi_r(p)\sim\mathrm{Gumbel}(0,1)\) is added to each candidate score. The corresponding soft assignment probability is
\begin{equation}
\pi_r(p)
=
\frac{
\exp\left(
\left[\ell_r(p)+\xi_r(p)\right]/T
\right)
}{
\displaystyle
\sum_{r'\in\{-,+\}}
\exp\left(
\left[\ell_{r'}(p)+\xi_{r'}(p)\right]/T
\right)
},
\label{eq:gumbel_assignment}
\end{equation}
where \(T\) denotes the annealing temperature. The relaxed latent value is then obtained as
\begin{equation}
\widetilde{I}(p)
=
\pi_{-}(p)I^{-}(p)
+
\pi_{+}(p)I^{+}(p).
\label{eq:sga_rounding}
\end{equation}
As \(T\) decreases, the soft assignment becomes increasingly concentrated on one of the two integer candidates, allowing \(\widetilde{I}(p)\) to better approximate the discrete floor-or-ceil decision in the later stage of refinement.

The relaxed latent grids are optimized using the rate-distortion objective
\begin{equation}
\mathcal{L}_{\mathrm{ref}}
=
\mathbb{E}_{\boldsymbol{\xi}}
\left[
D\left(\mathbf{x},f_{\theta}(\widetilde{I}_{\boldsymbol{\xi}})\right)
+
\lambda R
\right],
\label{eq:sga_objective}
\end{equation}
where \(\boldsymbol{\xi}\) denotes the sampled Gumbel variables, \(D\) is the mean squared error, and \(R\) denotes the stage-dependent rate estimate described below. Here, \(\lambda\) denotes the Lagrange multiplier that determines the operating rate point.

During the first quarter of refinement, the rate is evaluated directly on the relaxed latent values to preserve smooth gradients. During the remaining iterations, it is replaced by the expected coding cost of the two integer candidates:
\begin{equation}
R_{\mathrm{exp}}
=
-\sum_p\sum_{r\in\{-,+\}}
\pi_r(p)
\log_2
P\left(
I^r(p)\mid
\mathrm{context}_{\widetilde{I}}(p)
\right),
\label{eq:expected_integer_rate}
\end{equation}
where \(P(*)\) denotes the discrete probability mass of the corresponding integer symbol derived from the distribution predicted by LC3EM. \(p\) indexes the latent elements across all grids. This two-stage strategy progressively aligns the relaxed optimization objective with the integer-valued coding process.

In practice, the expectation in \eqref{eq:sga_objective} is approximated using three independently sampled realizations per update. We perform 2,000 Adam updates, corresponding to only 2\% additional optimization iterations. We set \(\alpha=1.2\), geometrically anneal \(T\) from \(0.3\) to \(0.08\), and reduce the learning rate from \(5\times10^{-4}\) to \(1\times10^{-5}\) using cosine annealing. The refined latent grids are finally hard-rounded for entropy coding.

\begin{figure*}[t]
\centering
\subfloat[SIQAD\label{fig:rd_siqad}]{%
    \includegraphics[width=0.48\textwidth]{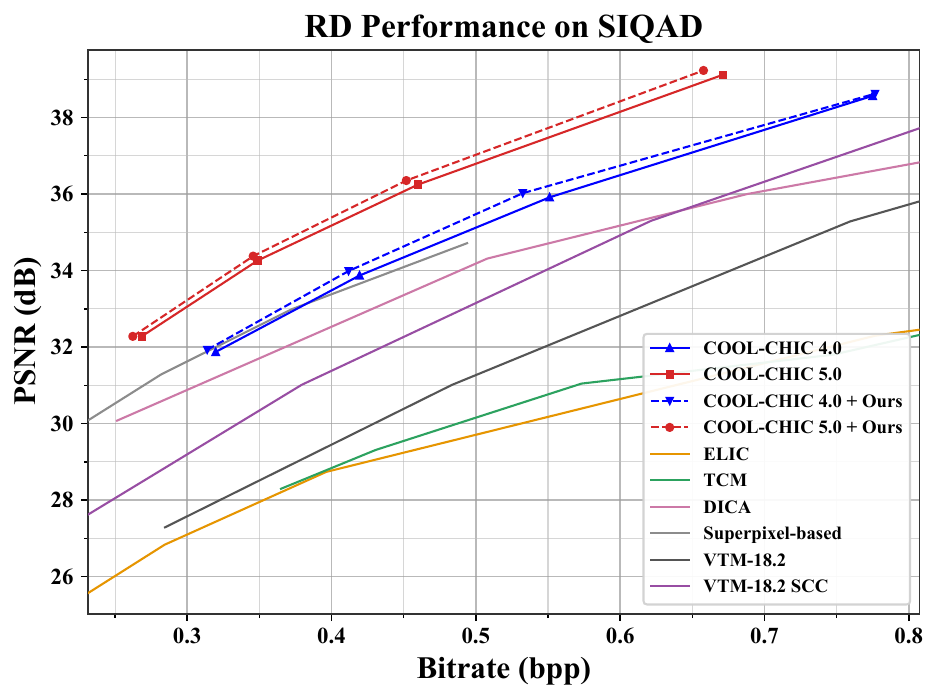}}
\hfill
\subfloat[API\label{fig:rd_api}]{%
    \includegraphics[width=0.48\textwidth]{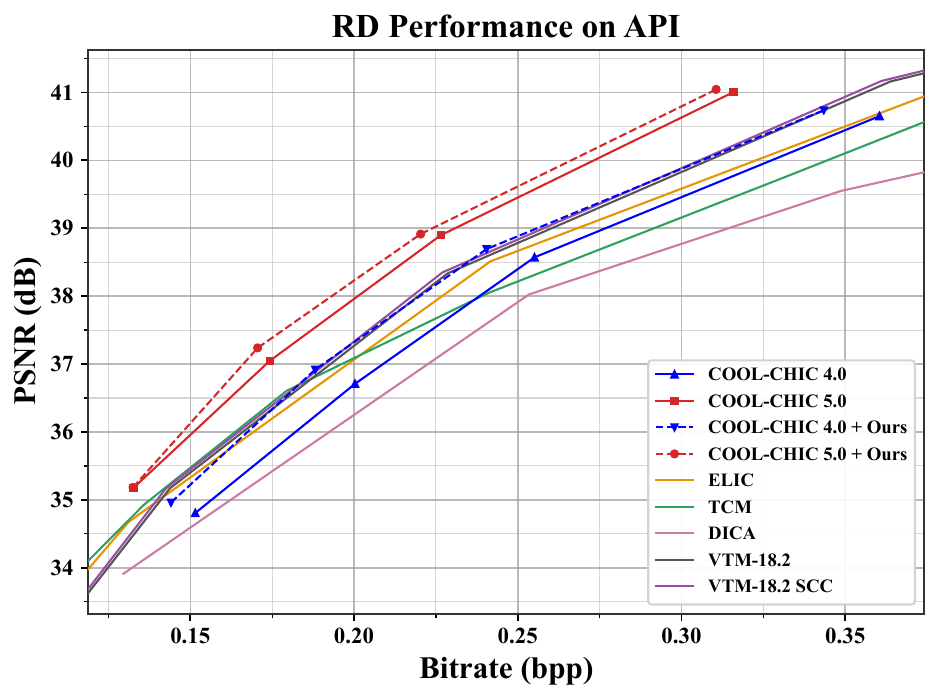}}

\medskip

\subfloat[SCID\label{fig:rd_scid}]{%
    \includegraphics[width=0.48\textwidth]{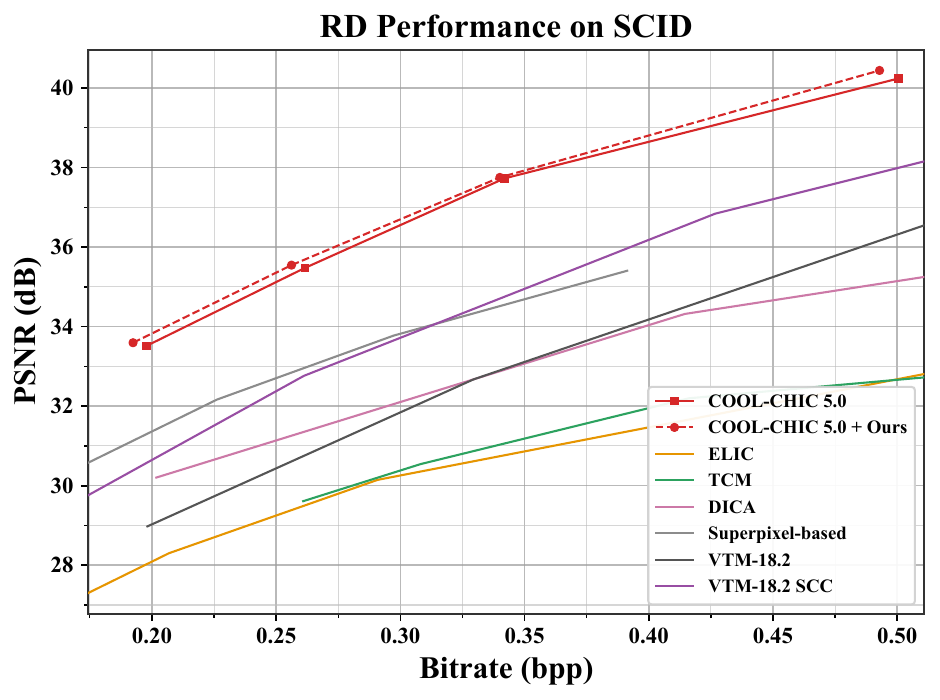}}
\hfill
\subfloat[Kodak\label{fig:rd_kodak}]{%
    \includegraphics[width=0.48\textwidth]{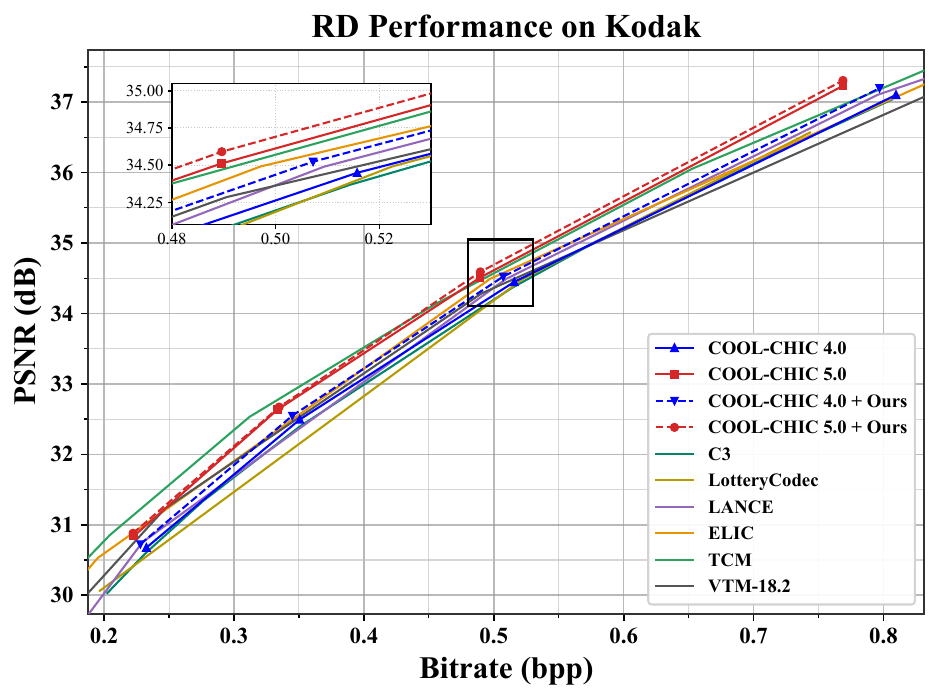}}
\caption{RD performance comparison on (a) SIQAD, (b) API, (c) SCID, and (d) Kodak. Methods are shown only on datasets for which results are available. The blue and red solid curves denote the results of the corresponding COOL-CHIC backbones, while the dashed curves of the same colors represent the performance after integrating the proposed method.}
\label{fig:rd_comparison}
\end{figure*}

\section{experiments and analysis}
\label{sec5}

\subsection{Experimental setup}

The experiments are conducted on both COOL-CHIC 4.0 and the latest COOL-CHIC 5.0 to validate the effectiveness and generality of the proposed method. The ``hop'' decoding configuration and the ``slow\_100k'' encoding configuration are adopted to achieve the best RD performance. All training hyperparameters are kept identical to those of the baselines, and both the baselines and the proposed method are retrained to obtain complete RD curves. To evaluate the performance across images with different characteristics, we conduct experiments on four benchmark datasets covering three content categories: one natural image dataset, Kodak~\cite{franzen_kodak_truecolor}; two screen content image datasets, SCID~\cite{ni2017scid} and SIQAD~\cite{yang2015perceptual}; and one anime image dataset, API~\cite{wang2024apisr}.



\noindent\textbf{Kodak:} The Kodak dataset contains 24 natural RGB images with a resolution of $768 \times 512$.

\noindent\textbf{SCID:} The SCID dataset contains 40 high-quality RGB screen content images of $1280 \times 720$, covering diverse content such as text, application interfaces, and games.

\noindent\textbf{SIQAD:} The SIQAD dataset contains 20 high-quality RGB screen content images with varying spatial resolutions, mainly mixing text and natural content.

\noindent\textbf{API:} The API dataset contains thousands of anime images. For the main evaluation, we use the first 10 RGB images according to the original dataset ordering. The exact image identifiers are provided in the accompanying repository for reproducibility.


\subsection{RD performance}
\label{RD}
In this section, the RD performance of the proposed method is evaluated against several advanced codecs, including overfitting-based image compression frameworks such as COOL-CHIC~4.0, COOL-CHIC~5.0~\cite{ladune2026cool}, C3~\cite{kim2024c3}, LotteryCodec~\cite{wu2025lotterycodec} and LANCE~\cite{benjak2026lance}. We also compare against autoencoder-based compression methods, including advanced models developed for natural images, such as ELIC~\cite{he2022elic} and TCM~\cite{liu2023learned}, as well as methods specifically designed for screen content images, including DICA~\cite{wang2025text} and the superpixel-based method~\cite{shen2024efficient}. In addition, the traditional codec VTM-18.2 is included as a reference. For screen content and anime images, in addition to the standard All-Intra configuration, we further evaluate VTM using the {classF.cfg} configuration, denoted as VTM-18.2-SCC, to better reflect its coding performance on computer-generated images. The corresponding RD results are presented in Figure~\ref{fig:rd_comparison}.

Integrating the proposed method into the COOL-CHIC series consistently improves RD performance across different datasets and backbones. When the state-of-the-art overfitting-based codec COOL-CHIC~5.0 is adopted as the backbone, the proposed method achieves BD-rate gains of -2.88\%, -2.39\%, -1.06\% and -3.15\% on the SIQAD, SCID, Kodak and API datasets, respectively. We further evaluate the proposed method using COOL-CHIC~4.0 as the backbone on the SIQAD, Kodak and API datasets, where additional BD-rate gains of -3.43\%, -2.94\%  and -7.69\% are achieved, respectively. These consistent improvements demonstrate the effectiveness of the proposed method and its generalizability across different generations of the COOL-CHIC framework.

Compared with autoencoder-based and traditional codecs, the proposed method built upon COOL-CHIC~5.0 achieves competitive RD performance on the natural image Kodak dataset, while substantially outperforming the compared methods on the computer-generated SIQAD, SCID, and API datasets. Notably, the performance gains brought by the proposed method are more pronounced on computer-generated content, including screen content and anime images, than on natural images. This observation indicates a clear content-dependent behavior of the proposed method, which is further analyzed in Section~\ref{content_dependent_behavior}.

\subsection{Ablation study}
\label{ablation}

To assess the contribution of each component in the proposed method, we conduct a series of ablation experiments on the Kodak and API datasets using COOL-CHIC~4.0 as the backbone. The ablation study examines the individual effects of NLEM, MECMS, and the latent rounding refinement stage. The corresponding results are summarized in Table~\ref{tab:ablation}.

\noindent\textbf{MECMS:} MECMS introduces a minimum-entropy-inspired mechanism to fuse the NLEM and MLP-based local predictors at the distribution level. To evaluate its effectiveness, we construct a degraded variant in which the two predictions are combined using simple linear interpolation, i.e., $\mu^\ast = w \mu_1 + (1 - w)\mu_2$, while the scale parameter remains unchanged. A comparison between the first and second configurations in Table~\ref{tab:ablation} shows that MECMS achieves additional BD-rate gains of -1.10\% on the Kodak dataset and -4.64\% on the API dataset. These results demonstrate that distribution-level fusion, which jointly adjusts the predicted mean and its uncertainty, is more effective than directly interpolating the predicted values, thereby validating the effectiveness of the minimum-entropy-inspired fusion strategy adopted in MECMS.

\begin{table}[t]
\centering
\caption{Ablation study of the proposed method using COOL-CHIC~4.0 as the backbone.}
\label{tab:ablation}
\renewcommand{\arraystretch}{1.05}
\setlength{\tabcolsep}{7pt}
\begin{tabular}{lcccc}
\toprule
\multirow{2}{*}{Dataset}
& \multicolumn{2}{c}{LC3EM}
& \multirow{2}{*}{\raisebox{-0.5ex}{\shortstack{Training\\Refinement}}}
& \multirow{2}{*}{BD-Rate (\%)} \\
\cmidrule(lr){2-3}
& NLEM & MECMS & & \\
\midrule

\multirow{3}{*}{Kodak}
& \checkmark & $\times$ & $\times$ & -0.65 \\
& \checkmark & \checkmark & $\times$ & -1.75 \\
& \checkmark & \checkmark & \checkmark & \textbf{-2.94} \\

\midrule

\multirow{3}{*}{API}
& \checkmark & $\times$ & $\times$ & -1.65 \\
& \checkmark & \checkmark & $\times$ & -6.29 \\
& \checkmark & \checkmark & \checkmark & \textbf{-7.69} \\

\bottomrule
\end{tabular}
\end{table}

\noindent\textbf{Training refinement:} The relaxation noise introduced during optimization creates a mismatch between the training process and the actual coding process based on rounding quantization. Consequently, latent features located near rounding boundaries may suffer noticeable performance degradation due to suboptimal rounding decisions during inference. To alleviate this mismatch, an additional latent rounding refinement stage is introduced to directly optimize the quantized latent representation. As shown in Table~\ref{tab:ablation}, although this iterative refinement accounts for less than 2\% of the overall optimization iterations, it consistently improves the RD performance. Specifically, it achieves an additional BD-rate gain of -1.19\% on the Kodak dataset and a further gain of -1.40\% on the API dataset, thereby validating the effectiveness of the proposed training refinement strategy.

\begin{table}[t]
\centering
\setlength{\tabcolsep}{7pt}
\caption{Effect of different activation functions for constraining the range of $w$ in MECMS. Results are obtained on Kodak with COOL-CHIC~4.0 as the backbone at $\lambda=0.001$, without the additional latent rounding refinement; $D+\lambda R$ is reported after scaling by $10^3$.}
\begin{tabular}{lccc}
\toprule
Activation & PSNR (dB) & Rate (bpp) & $D+\lambda R$ \\
\midrule
$\bm{1.5\tanh(\cdot)_{+}}$ & 34.45 & \textbf{0.507} & \textbf{0.866} \\
$\tanh(\cdot)_{+}$         & 34.45 & 0.509 & 0.868 \\
Sigmoid                    & 34.46 & 0.510 & 0.868 \\
\bottomrule
\end{tabular}
\label{tab:LC3EM_activation}
\end{table}

Beyond the ablation of the overall components, we further analyze the extrapolative mechanism of the two prediction modes in MECMS, where the fusion weight $w$ is extended from $[0,1]$ to $[0,1.5]$ using $1.5\tanh(\cdot)_+$. For comparison, we also construct a non-extrapolative variant in which the fusion weights are constrained to $[0,1]$ using $\tanh(\cdot)_+$ and sigmoid functions. The comparison at $\lambda = 0.001$, reported in Table~\ref{tab:LC3EM_activation} shows that restricting $w$ to $[0,1]$ results in a clear performance degradation. We attribute this gain not only to the extrapolative capability itself, but also to the increased flexibility of the fusion mechanism, which allows the model to alleviate performance degradation by enlarging the scale parameter in regions where effective fusion is difficult.

\begin{figure*}[h]
    \centering
    \includegraphics[width=0.95\linewidth]{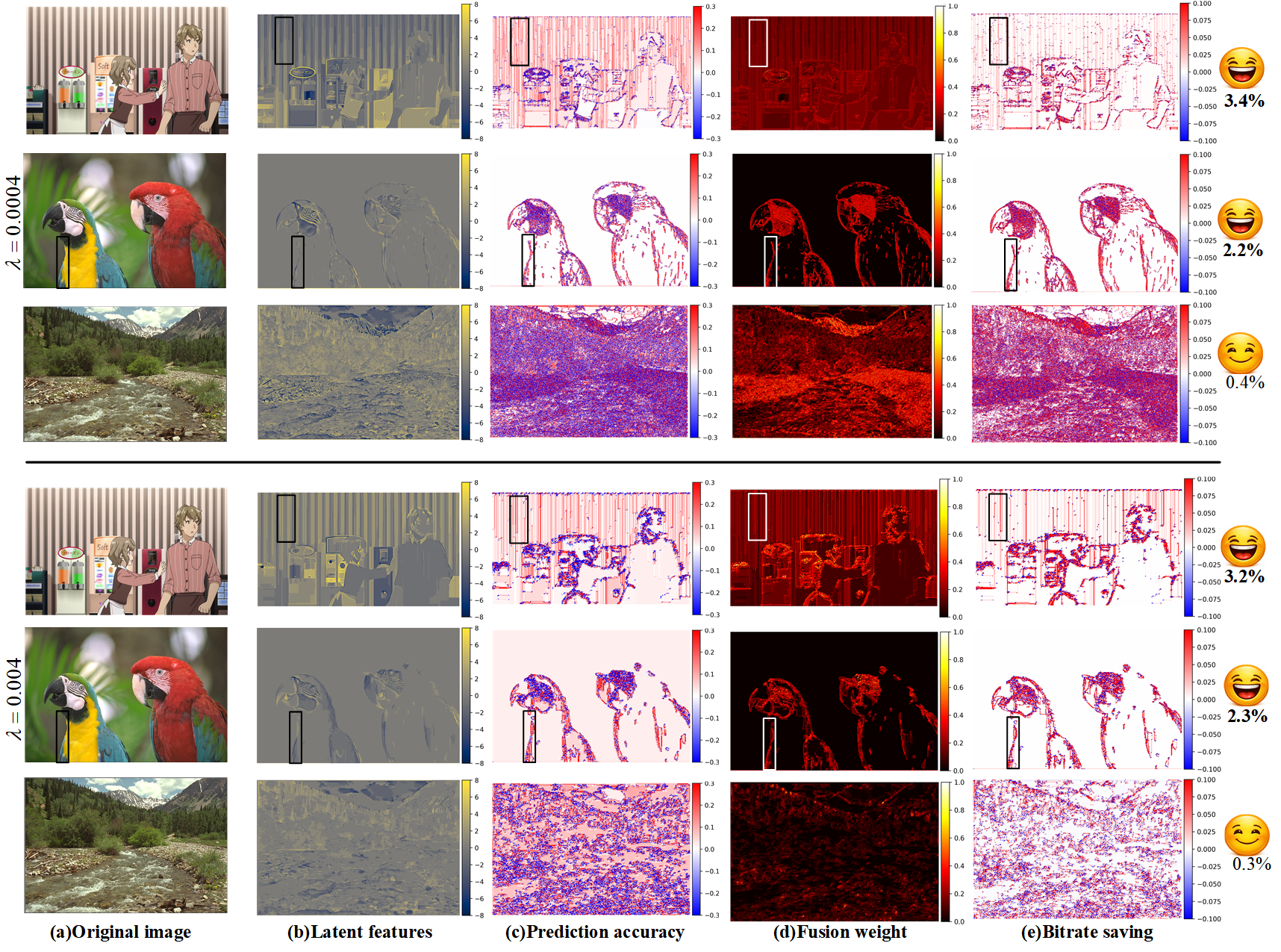}

    \caption{Visualization analysis under the high-bitrate setting ($\lambda = 0.0004$) and the low-bitrate setting ($\lambda = 0.004$). One anime image from the API dataset and two natural images from the Kodak dataset are selected for comparison. (c) shows $\lvert I - \mu_{\text{MLP}} \rvert - \lvert I - \mu_{\text{NLEM}} \rvert$, where positive values indicate more accurate predictions from NLEM. (d) presents the fusion weight $(1-w)$, where values closer to 1 indicate a larger contribution from NLEM, while 0 indicates no contribution (cases with $w>1$ are excluded). (e) illustrates the bitrate savings, defined as $R(\mu_1,b_1)-R(\mu^{*},b^{*})$, where positive values indicate bitrate reduction and negative values indicate bitrate increase. The last column reports the percentage improvement in RD performance ($D+\lambda R$) achieved by LC3EM for each image.}
    \label{fig:delta_function}
\end{figure*}

\subsection{LC3EM performance analysis}

This section focuses on analyzing the performance of LC3EM from three perspectives: algorithm design, content-dependent behavior and bitstream composition. The latent rounding refinement stage is disabled in this section to control variables and ensure a fair analysis.

\subsubsection{Effective algorithm design}

The performance gains of LC3EM primarily stem from two aspects: the superior capability of NLEM over the MLP-based predictor in COOL-CHIC for modeling long-range directional redundancies, and the effective fusion mechanism enabled by MECMS. Together, these components form the core algorithmic design underlying the improved performance of LC3EM.

In terms of prediction accuracy, Figure~\ref{fig:delta_function}(c) shows that NLEM is particularly effective for directional structures, especially in edge regions and line-like textures. Such regions exhibit long-range directional patterns that are difficult for the local MLP-based predictor to capture due to its limited receptive field. In contrast, NLEM exploits recurring structural patterns through template matching and least-squares estimation, thereby providing more accurate predictions. Regarding the fusion behavior, Figure~\ref{fig:delta_function}(d) visualizes the fusion weights determined by MECMS. Edge regions generally appear brighter, indicating that larger values of \(1-w\), corresponding to a larger NLEM contribution, are observed in regions where it provides more accurate predictions. This adaptive fusion further improves the accuracy of entropy estimation. Correspondingly, Figure~\ref{fig:delta_function}(e) illustrates the bitrate savings achieved by the fusion mechanism. The bitrate reduction is mainly concentrated in regions where NLEM performs better, further validating the effectiveness of the proposed design. The rectangular boxes highlight representative regions in which accurate NLEM predictions, larger NLEM contributions, and substantial bitrate savings are consistently observed.

\subsubsection{Content-dependent behavior}
\label{content_dependent_behavior}

As a deterministic prediction mechanism, NLEM mainly benefits from its ability to effectively exploit long-range contexts and efficiently model directional structural redundancy, which is difficult for the MLP-based predictor with a limited receptive field to capture. Consequently, LC3EM naturally exhibits content-dependent behavior, providing larger performance gains for images in which such characteristics are more prominent. Compared with natural images, computer-generated content, including screen content and anime images, generally contains more pronounced sharp edges, repetitive patterns, and regular structures~\cite{adhuran2020parameter,wang2024dscic}, making its structural redundancy more amenable to the long-range directional prediction provided by NLEM. As reported in Section~\ref{RD}, when integrated into the COOL-CHIC series, the proposed method achieves larger BD-rate gains on the computer-generated SIQAD, SCID, and API datasets than on the natural-image Kodak dataset. Figure~\ref{fig:delta_function} provides consistent visual evidence: compared with natural images, the anime-image cases (the first and fourth rows) exhibit more pronounced advantages of NLEM, leading to larger compression gains.

A clear content-dependent tendency can also be observed within natural images in Figure~\ref{fig:delta_function}. The comparison of two representative images at two bitrate points clearly reveals the variation in performance gains across different image contents. For the natural image shown in the second row, the feather textures of the bird are highly favorable to NLEM. Since these directional structures occupy a large proportion of the image content, substantial RD performance improvements are achieved, reaching $2.2\%$ at $\lambda=0.0004$. In contrast, when directional structural redundancy is less prominent, the prediction difference between NLEM and the learned local predictor becomes less significant and exhibits no clear spatial regularity. As shown in the image containing rivers and trees in the third row of Figure~\ref{fig:delta_function}, it becomes more challenging for the model to identify the optimal prediction mode for different regions, resulting in a limited gain of only $0.4\%$ at $\lambda=0.0004$. (Here, the reported percentage improvement is measured by the relative reduction in the normalized $D+\lambda R$ objective and should not be interpreted as a BD-rate gain.)

\subsubsection{Bitstream composition}

For existing coordinate-based overfitting codecs, a fundamental trade-off exists in entropy model design. Although a powerful entropy model can compress the latent representation more efficiently, it typically incurs a higher parameter transmission overhead, making direct model scaling ineffective in terms of overall coding performance. In contrast, the proposed method enhances the capability of the entropy model with almost no additional transmission overhead, which constitutes a key factor behind the performance gains of LC3EM.

\begin{table}[t]
\centering
\caption{Bitstream composition (encoding-side estimation) for different methods on Kodak 15 with $\lambda=0.001$. $D+\lambda R$ is reported after scaling by $10^3$.}
\setlength{\tabcolsep}{1.8pt}
\begin{tabular}{lcccc}
\toprule
Method & $D+\lambda R$ & Network bpp & Latent bpp & Total bpp \\
\midrule
COOL-CHIC~4.0
& 0.6138 & \textbf{0.0309} & 0.2792 & 0.3101 \\
COOL-CHIC~4.0 (32 ctxs)
& 0.6165 & 0.0615 & \textbf{0.2550} & 0.3165 \\
COOL-CHIC~4.0+LC3EM
& \textbf{0.6000} & 0.0312 & 0.2674 & \textbf{0.2986} \\
\bottomrule
\end{tabular}
\label{tab:bit_composition}
\end{table}

To further illustrate the core idea of LC3EM, we construct an enhanced variant of COOL-CHIC~4.0 by extending the context input of its entropy model from the original 16 elements to 32 elements, denoted as COOL-CHIC~4.0 (32 ctxs). Table~\ref{tab:bit_composition} compares the RD performance and bitstream composition of the baseline COOL-CHIC~4.0, its 32-context variant, and the proposed LC3EM-enhanced method. The results show that enlarging the context and network capacity indeed improves entropy modeling, reducing the latent bitrate from 0.2792 to 0.2550~bpp. However, this gain comes at a substantial cost: the network bitrate nearly doubles from 0.0309 to 0.0615~bpp, resulting in a higher total bitrate of 0.3165~bpp and even a degraded overall RD performance.  In contrast, since NLEM is a handcrafted mode, it introduces no additional parameter transmission overhead. As a result, LC3EM increases the network bitrate by only 0.0003~bpp, corresponding solely to the parameters of an additional fusion layer, while achieving more accurate entropy estimation of the latents and, consequently, better overall coding performance.

\subsection{Decoding complexity analysis}

This section evaluates the computational complexity of the proposed method and investigates the trade-off between computational cost and coding performance. The main overhead arises from NLEM, particularly the construction and solution of the local least-squares normal equations. Increasing the number of contextual equations does not introduce additional transmission overhead, since the required context samples are already available at the decoder, but it increases the computational cost. Table~\ref{tab:post_training_LC3EM} reports the coding performance and decoder-side complexity on Kodak and API, using COOL-CHIC~4.0 as the backbone. The full method introduces approximately 1188 additional MACs/pixel.

\begin{table}[t]
\centering
\caption{BD-rate and decoding complexity of the proposed method and its pruning variants on the Kodak and API datasets using COOL-CHIC~4.0 as the backbone. BD-rate is measured relative to the corresponding COOL-CHIC~4.0 anchor.}
\label{tab:post_training_LC3EM}
\renewcommand{\arraystretch}{1.0}
\setlength{\tabcolsep}{7pt}
\begin{tabular}{llcc}
\toprule
Dataset
& Method
& \makecell{BD-rate\\(\%)}
& \makecell{Complexity\\(MACs/pixel)} \\
\midrule

\multirow{5}{*}{Kodak}
& Anchor                      &  0.00 & 1433 \\
& Ours (full)                 & -2.94 & 2621 \\
& Ours ($s=24$)               & -2.59 & 2194 \\
& Ours ($\tau=0.1$)           & -2.82 & 1910 \\
& Ours ($s=24,\tau=0.1$)      & -2.48 & 1743 \\

\midrule

\multirow{5}{*}{API}
& Anchor                      &  0.00 & 1433 \\
& Ours (full)                 & -7.69 & 2621 \\
& Ours ($s=24$)               & -7.16 & 2194 \\
& Ours ($\tau=0.1$)           & -7.40 & 2214 \\
& Ours ($s=24,\tau=0.1$)      & -6.91 & 1936 \\

\bottomrule
\end{tabular}
\end{table}

Two post-training pruning strategies are considered to reduce decoding complexity with only a minor loss in coding performance. The first strategy reduces the number of equations used by NLEM, decreasing the equation count from 40 to \(s=24\). The second strategy skips NLEM at positions satisfying \(|w-1|\leq\tau\), where the contribution of NLEM to the fused prediction is limited. With \(\tau=0.1\), approximately 60.98\% and 34.90\% of NLEM evaluations are skipped on the Kodak and API datasets, respectively. The two strategies can also be combined. On Kodak, the decoding complexity is reduced from 2621 to 1743 MACs/pixel, leaving an overhead of only 310 MACs/pixel over the anchor while retaining a \(-2.48\%\) BD-rate gain. On API, the decoding complexity decreases from 2621 to 1936 MACs/pixel while retaining a \(-6.91\%\) BD-rate gain.

\section{Conclusion}
\label{sec6}
This paper proposes a Long-Range Context Extrapolation Enhanced Entropy Model for coordinate-based overfitting image codecs, aiming to improve the efficiency of entropy modeling in exploiting long-range contextual information and directional structures. Furthermore, an auxiliary rounding refinement stage is introduced to improve compression performance. The effectiveness of the method is demonstrated by integrating it into the state-of-the-art overfitting image codec COOL-CHIC~5.0 and its predecessor, COOL-CHIC~4.0. More broadly, this work demonstrates the potential of coding mode selection for scenarios where the entropy model size is constrained, by introducing customizable handcrafted prediction modes that can be adaptively selected for different regions. Beyond NLEM, future work will explore complementary prediction modes within the COOL-CHIC framework to better accommodate images with diverse characteristics.

\bibliographystyle{IEEEtran}
\bibliography{ref}

\end{document}